\documentclass[12pt]{article}
\usepackage{graphicx}
\begin{document}
\thispagestyle{empty}
\begin{center}

\vspace*{1cm} {\Large  \bf The Mirror Universes}

\bigskip
{\sc Miroslaw Kozlowski}$^{\rm a}$ and  {\sc Janina
Marciak-Kozlowska}

\bigskip
Institute of Electron Technology Al.~Lotników~32/46 02--668~Warsaw
Poland

 \hbox to 5cm{\hsize=5cm\vbox{\ \hrule}}\par $^{\rm
a}$~Author to whom correspondence should be
addressed:MiroslawKozlowski@aster.pl
\end{center}

\bigskip
\begin{abstract}
In this paper we investigate the structure of the Mirror
Universes. The two universes are coupled with transformation
$t\to-t$. It is shown that for Planck scale i.e. for $t\cong
T_{\rm Planck}=\left(\frac{\hbar G}{c^5}\right)^{1/2}$ the
oscillations of temperature of the universes are observed. For the
Mirror Universes the temperature fields are shifted in phase.

\end{abstract}

\newpage
\section{Introduction}

The physical phenomenon of gravity, described to a high degree of
accuracy by Isaac Newton's mathematics in 1687, has played a key
role in scientific understanding. However, in 1915 Einstein
created a major revolution in our scientific worldview. According
to Einstein  theory gravity plays a unique role in physics for
several reasons~\cite{1}. Most particularly, these are: (i)
gravity is the only physical quality that influences causal
relationships between space-time events, (ii) Gravitation force
has no local reality, as it can be eliminated by a change in
space-time coordinates; instead gravitational tidal effects
provide a curvature for the very space-time in which all other
particles and forces are contained. It follows from this that
gravity cannot be regarded as some kind of emergent phenomenon
secondary to other physical effects, but is a fundamental
component of physical reality.

According to modern accepted physical pictures, reality is rooted
in three-dimensional space and a one-dimensional time, combined
together into four dimensional space-time. As was stated in
Penrose-Hameroff model conscious events are embedded at the Planck
scale, and is gravity dependent~\cite{2}.

In this paper we study the influence of gravity on the thermal
phenomena at two-dimensional blisters (1D space, 1D time) in
Planck scale. We will consider the Mirror Universes, i.e. two set
of the Universes coupled by the transformation $t\to-t$.
It will be shown that the temperature of the Universes are shifted
in phase. For short times i.e. for Planck scale, the oscillations
of the Universe temperature are predicted.

In this paper we follow of idea of the repulsive gravity as the
source of the space-time expansion. We will study the influence of
the repulsive gravity $(G<0)$ on the temperature field in the
universe. To that aim we will apply the quantum hyperbolic heat
transfer equation (QHT) formulated in our earlier papers~\cite{3,
4}.

When substitution $G\rightarrow-G$ is performed in QHT the
Schr{\"o}dinger type equation is obtained for the temperature field. In
this paper the solution of QHT will be obtained. The resulting
temperature is a complex function of space and time. We argue that
because of the anthropic limitation of the observers it is quite
reasonable to assume ${\rm Im}T=0$. From this anthropic condition the
discretization of the space radius $R=[(4N\pi+3\pi)L_{\rm
P}]^{1/2}(ct)^{1/2}$, velocity of expansion
\mbox{$v=(\pi/4)^{1/2}((N+\frac34)/{M})^{1/2}c$ } and acceleration of
expansion $a=-\frac12(\pi/4)^{1/2}((N+\frac34)^{1/2}/M^{3/2})(c^7/(\hbar
G))^{1/2}$ are obtained.

\section{The model}
In papers~\cite{3,4} the quantum heat transport equation in a Planck Era
was formulated:
\begin{equation}
\tau\frac{\partial^2 T}{\partial t^2}+\frac{\partial T}{\partial
t}=\frac{\hbar}{M_{\rm P}}\nabla^2 T.\label{eq1}
\end{equation}
In equation~(\ref{eq1}) $\tau=((\hbar G)/c^5)^{1/2}$ is the relaxation
time, $M_{\rm P}=((\hbar c)/G)^{1/2}$ is the mass of the Planck particle,
$\hbar, c$ are the Planck constant and light velocity respectively and
$G$ is the gravitational constant. The crucial role played by gravity
(represented by $G$ in formula~(\ref{eq1})) in a Planck Era was
investigated in paper~\cite{4}.

For a long time the question whether, or not the fundamental constant of
nature $G$ vary with time has been a question of considerable interest.
Since P.~A.~M.~Dirac~\cite{5} suggested that the gravitational force may
be weakening with the expansion of the Universe, a variable $G$ is
expected in theories such as the Brans-Dicke scalar-tensor theory and its
extension~\cite{6,7}. Recently the problem of the varying $G$ received
renewed attention in the context of extended inflation
cosmology~\cite{8}.

It is now known, that the spin of a field (electromagnetic, gravity) is
related to the nature of the force: fields with odd-integer spins can
produce both attractive and repulsive forces; those with even-integer
spins such as scalar and tensor fields produce a purely attractive force.
Maxwell's electrodynamics, for instance can be described as a spin one
field. The force from this field is attractive between oppositely charged
particles and repulsive between similarly charged particles.

The integer spin particles in gravity theory are like the graviton,
mediators of forces and would generate the new effects. Both the
graviscalar and the graviphoton are expected to have the rest mass and so
their range will be finite rather than infinite. Moreover, the
graviscalar will produce only attraction, whereas the graviphoton effect
will depend on whether the interacting particles are alike or different.
Between matter and matter (or antimatter and antimatter) the graviphoton
will produce repulsion. The existence of repulsive gravity forces can to
some extent explains the early expansion of the Universe~\cite{5}.

In this paper we will describe the influence of the repulsion gravity on
the quantum thermal processes in the universe. To that aim we put in
equation~(\ref{eq1}) $G\rightarrow-G$. In that case the new equation is
obtained, viz.
\begin{equation}
i\hbar\frac{\partial T}{\partial
t}=\left(\frac{\hbar^3|G|}{c^5}\right)^{1/2}\frac{\partial^2 T}{\partial
t^2}-\left(\frac{\hbar^3|G|}{c}\right)^{1/2}\nabla^2 T.\label{eq2}
\end{equation}
For the investigation of the structure of equation~(\ref{eq2}) we put:
\begin{equation}
\frac{\hbar^2}{2m}=\left(\frac{\hbar^3|G|}{c}\right)^{1/2}\label{eq3}
\end{equation}
and obtains
$$
m=\frac12M_{\rm P}
$$
with new form of the equation~(\ref{eq2})
\begin{equation}
i\hbar\frac{\partial T}{\partial
t}=\left(\frac{\hbar^3|G|}{c^5}\right)^{1/2} \frac{\partial^2 T}{\partial
t^2}-\frac{\hbar^2}{2m}\nabla^2T.\label{eq4}
\end{equation}
Equation~(\ref{eq4}) is the quantum telegraph equation discussed in
paper~\cite{4}. To clarify the physical nature of the solution of
equation~(\ref{eq4}) we will discuss the diffusion approximation, {\it
i.e.}~we omit the second time derivative in equation~(\ref{eq4}) and
obtain
\begin{equation}
i\hbar\frac{\partial T}{\partial
t}=-\frac{\hbar^2}{2m}\nabla^2T.\label{eq5}
\end{equation}
Equation~(\ref{eq5}) is the Schr{\"o}dinger type equation for the
temperature field in a universes with $G<0$.

Both equation~(\ref{eq5}) and diffusion equation:
\begin{equation}
\frac{\partial T}{\partial t}=\frac{\hbar^2}{2m}\nabla^2T\label{eq6}
\end{equation}
are parabolic and require the same boundary and initial conditions in
order to be ``well posed''.

The diffusion equation~(\ref{eq6}) has the propagator~\cite{10}:
\begin{equation}
T_D(\vec{R}, \Theta)=\frac{1}{(4\pi D\Theta)^{3/2}} \exp
\left[-\frac{R^2}{2\pi \hbar \Theta}\right],\label{eq7}
\end{equation}
where
$$
\vec{R}=\vec{r}-\vec{r'}, \qquad \Theta=t-t'.
$$
For equation~(\ref{eq5}) the propagator is:
\begin{equation}
T_s(\vec{R}, \Theta)=\left(\frac{M_{\rm P}}{2\pi\hbar\Theta}\right)^{3/2}
\exp\left[-\frac{3\pi i}{4}\right]\cdot \exp\left[\frac{iM_{\rm
P}R^2}{2\pi\hbar\Theta}\right]\label{eq8}
\end{equation}
with initial condition $T_s(\vec{R}, 0)=\delta(\vec{R})$.

In equation~(\ref{eq8}) $T_s(\vec{R}, \Theta)$ is the complex function of
$\vec{R}$ and $\Theta$. For anthropic observers only the real part of $T$
is detectable, so in our description of universe we put:
\begin{equation}
{\rm Im}T(\vec{R}, \Theta)=0.\label{eq9}
\end{equation}
The condition~(\ref{eq9}) can be written as (bearing in mind
formula~(\ref{eq8})):
\begin{equation}
\sin\left[-\frac{3\pi}{4}+\left(\frac{R}{L_{\rm
P}}\right)^2\frac{1}{4\widetilde{\Theta}} \right]=0,\label{eq10}
\end{equation}
where $L_{\rm P}=\tau_{\rm P}c$ and $\widetilde{\Theta}=\Theta/\tau_{\rm
P}$. Formula~(\ref{eq10}) describes the discretization of~$R$
\begin{eqnarray}
R_N&=&[(4N\pi+3\pi)L_{\rm P}]^{1/2}(tc)^{1/2},\label{eq11}\\
N&=&0, 1, 2, 3\ldots\nonumber
\end{eqnarray}
In fact from formula~(\ref{eq11}) the Hubble law can be derived
\begin{equation}
\frac{\dot{R_N}}{R_N}=H=\frac{1}{2t}, \qquad {\rm independent \; of} \;
N.\label{eq12}
\end{equation}
In the subsequent we will consider $R$~(\ref{eq11}), as the space-time
radius of the $N-$~universe with ``atomic unit'' of space $L_{\rm P}$.

It is well known that idea of discrete structure of time can be applied
to the ``flow'' of time. The idea that time has ``atomic'' structure or
is not infinitely divisible, has only recently come to the fore as a
daring and sophisticated hypothetical concomitant of recent
investigations in the physics elementary particles and astrophysics. Yet
in the Middle Ages the atomicity of time was maintained by various
thinkers, notably by Maimonides~\cite{11}. In the most celebrated of his
works: {\it The Guide for perplexed} he wrote: {\it Time is composed of
time-atoms, i.e. of many parts, which on account of their short duration
cannot be divided.} The theory of Maimonides was also held by
Descartes~\cite{12}.

The shortest unit of time, atom of time is named {\it chronon}~\cite{13}.
Modern speculations concerning the {\it chronon} have often be related to
the idea of the smallest natural length is $L_{\rm P}$. If this is
divided by velocity of light it gives the Planck time $\tau_{\rm
P}=10^{-43}$~s, {\it i.e. the chronon} is equal $\tau_{\rm P}$. In that
case the time $t$ can be defined as
\begin{equation}
t=M\tau_{\rm P}, \qquad M=0, 1, 2, \dots\label{eq13}
\end{equation}
Considering formulae~(\ref{eq8}) and (\ref{eq13}) the space-time radius
can be written as
\begin{equation}
R(M, N)=(\pi)^{1/2}M^{1/2}\left(N+\frac34\right)^{1/2}L_{\rm P}, \qquad
M, N=0, 1, 2, 3, \ldots\label{eq14}
\end{equation}
Formula~(\ref{eq14}) describes the discrete structure of space-time. As
the $R(M, N)$ is time dependent, we can calculate the velocity, $v={\rm
d}R/{\rm d}t$, {\it i.e.} the velocity of the expansion of space-time
\begin{equation}
v=\left(\frac{\pi}{4}\right)^{1/2}
\left(\frac{N+3/4}{M}\right)^{1/2}c,\label{eq15}
\end{equation}
where $c$ is the light velocity. We define the acceleration of the
expansion of the space-time
\begin{equation}
a=\frac{{\rm d} v}{{\rm d} t}=-\frac12\left(\frac{\pi}{4}\right)^{1/2}
\frac{(N+3/4)^{1/2}}{M^{3/2}}\frac{c}{\tau_{\rm P}}.\label{eq16}
\end{equation}
Considering formula~(\ref{eq16}) it is quite natural to define Planck
acceleration:
\begin{equation}
A_{\rm P}=\frac{c}{\tau_{\rm P}}=\left(\frac{c^7}{\hbar
G}\right)^{1/2}=10^{51}\; {\rm ms}^{-2}\label{eq17}
\end{equation}
and formula~(\ref{eq16}) can be written as
\begin{equation}
a=-\frac12\left(\frac{\pi}{4}\right)^{1/2}
\frac{\left(N+3/4\right)^{1/2}}{M^{3/2}}\left(\frac{c^7}{\hbar
G}\right)^{1/2}.\label{eq18}
\end{equation}

\begin{table}[t]
\begin{center}
\caption{\it Radius, velocity and acceleration for N, M-universes}
\vspace{0.3cm}
\begin{tabular}{|c|l|c|l|}
\hline
&&&\\
N, M&R[{\rm m}]&v[{\rm m/s}]&a[{\rm m/s}$^2$]\\
&&&\\ \hline
&&&\\
$10^{20}$&$1.77 \cdot 10^{-15}$&$2.66 \cdot 10^8$&$-1.32 \cdot 10^{31}$\\
&&&\\ \hline
&&&\\
$10^{60}$&$1.77\cdot 10^{25\,(*)}$&$2.66 \cdot10^{8\,(*)}$&$-1.32
\cdot 10^{-10\,(**)}$\\
&&&\\ \hline
&&&\\
$10^{80}$&$1.77 \cdot 10^{45}$&$2.66 \cdot 10^8$&$-1.32 \cdot 10^{-29}$\\
&&&\\ \hline
\end{tabular}

\vspace{0.2cm}
\parbox{12cm}{
$^{(*)}$Spergel~D.~N.~{\it at~al.}~\cite{16};\\
$^{(**)}$Anderson~J.~D.~{\it at~al.}~\cite{17} Radio metric data
from Pionier~10/11, Galileo and Ulysses Data indicate and apparent
anomalous, constant, acceleration acting on the spacecraft with a
magnitude $\sim 8.5 \cdot 10^{-10} \;{\rm m/s}^2$.}
\end{center}
\end{table}

In table~I the numerical values for $ R, v$ and $a$ are presented. It is
quite interesting that for $N, M\rightarrow\infty$ the expansion
velocity\\
 $v<c$ in complete accord with relativistic description.
Moreover for $N, M\gg1$ the $v$ is relatively constant $v\sim 0.88 c$.
From formulae~(\ref{eq11}) and (\ref{eq15}) the Hubble parameter $H$, and
the age of our Universe can be calculated
\begin{eqnarray}
v&=&HR, \qquad H=\frac{1}{2M\tau_{\rm P}}=5\cdot10^{-18} \; {\rm s}^{-1},\nonumber \\
T&=&2M\tau_{\rm P}=2\cdot 10^{17} {\rm s}\sim10^{10}\; {\rm
years},\label{eq19}
\end{eqnarray}
which is in quite good agreement with recent measurement~\cite{15, 16,
17}.

In Fig.~1a,b we present the behaviour of $T_S(\frac{R}{R_{\rm
Planck}}, \, \frac{t}{T_{\rm Planck}})$ for positive as well
negative values of $R$ and $T$. For small values of $R$, $t$ (in
comparison to $R_{\rm Planck}$ and $T_{\rm Planck}$ one can
observe the vibrating structure of $T_S$ (formula~(\ref{eq8})) for
positive as well negative, $R$ and $t$ values. For the mirror
universe $t\to-t$, $R\to-R$ (in 1D case) the "phase shift" is
observed.
\newpage
\begin{figure}[h]
\centering\includegraphics[width=12 cm]{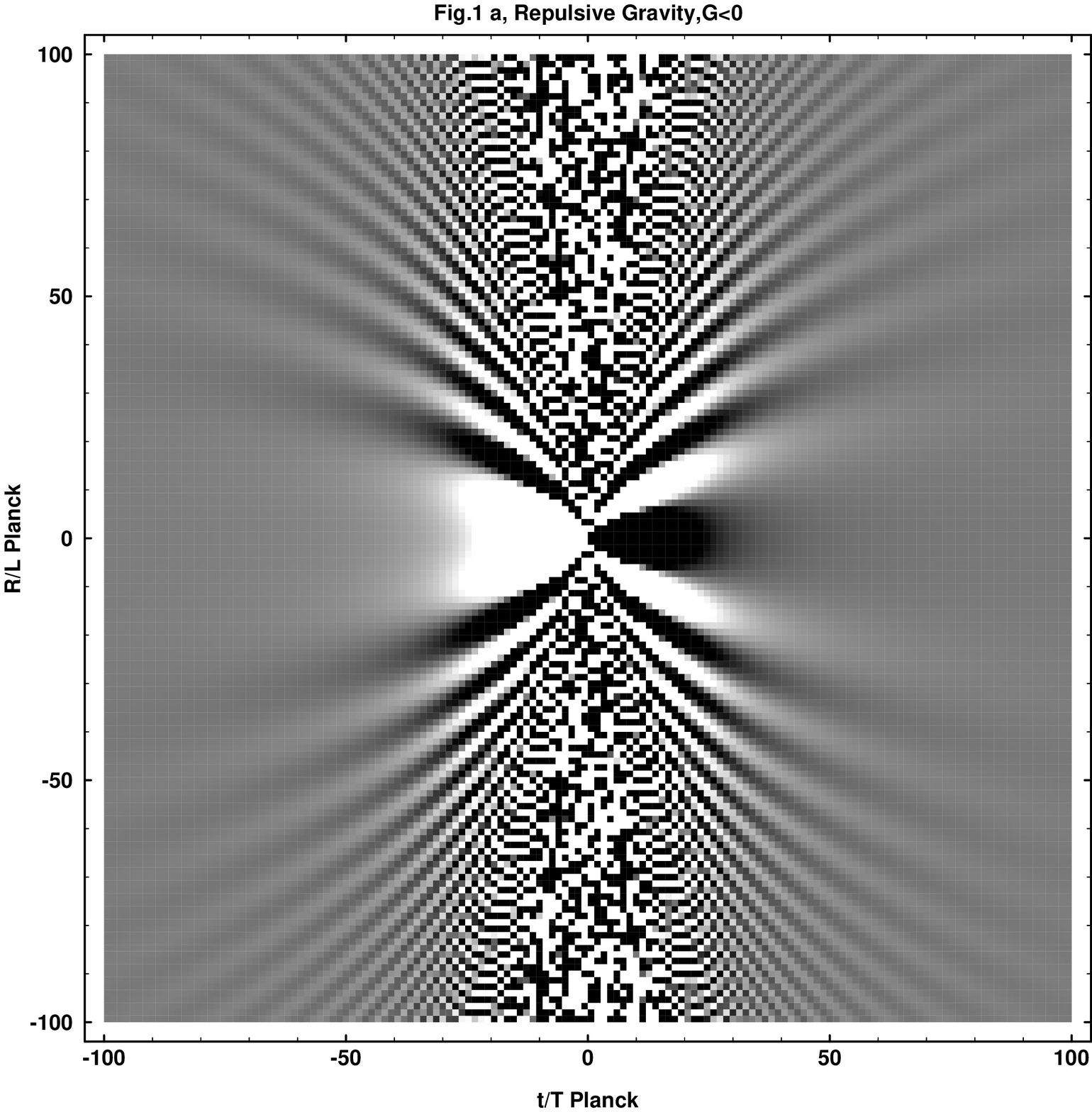}
\caption{The
temperature fields for Planck scales, (a) for $\frac{t}{T_{\rm
Planck}} (-100, 100)$}
\end{figure}
\begin{figure}
\centering\includegraphics[width=12 cm]{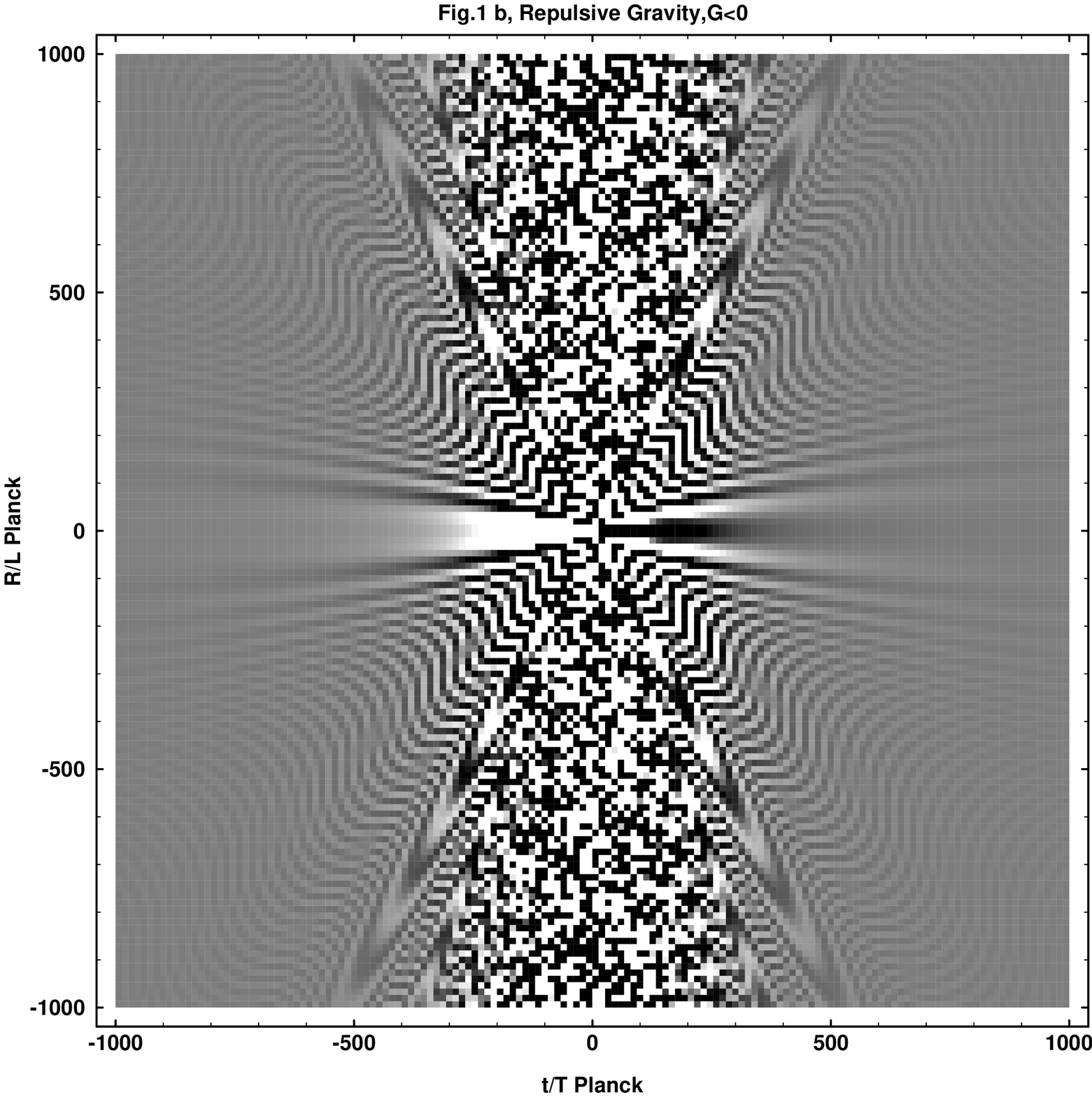} \caption{The
temperature fields for Planck scales,(b) for $\frac{t}{T_{\rm Planck}} (-1000,
1000)$}
\end{figure}

\end{document}